\documentclass[runningheads]{llncs}
\usepackage{graphicx}
\usepackage{booktabs}
\usepackage{tabularx}
\usepackage{hyperref}
\usepackage{amsmath}
\usepackage{adjustbox}
\usepackage{multirow}
\usepackage[table]{xcolor}
\usepackage{arydshln}
\usepackage{bbm}
\usepackage{caption}
\usepackage{subcaption}
\usepackage{amssymb}
\usepackage{marvosym}

\usepackage{array}
\usepackage{makecell}

\newif\ifworkinprogress
\workinprogresstrue

\ifworkinprogress
    \newcommand{\ms}[1]{\textcolor{darkgreen}{{[MS] #1}}}
    \newcommand{\dk}[1]{\textcolor{blue}{{[DK] #1}}}
    \newcommand{\pmu}[1]{\textcolor{teal}{{[PM] #1}}}
    \newcommand{\el}[1]{\textcolor{orange}{{[EL] #1}}}
\else
    \newcommand{\ms}[1]{}
    \newcommand{\dk}[1]{}
    \newcommand{\pmu}[1]{}
    \newcommand{\el}[1]{}
\fi

\newcommand{\movielens}{\emph{MovieLens 1M}}
\newcommand{\lfm}{\emph{LastFM User Groups}}
\newcommand{\amazon}{\emph{Amazon Grocery \& Gourmet}}
\newcommand{\recall}{\emph{Recall}}

\newcommand{\enmf}{\emph{ENMF}}
\newcommand{\lightgcn}{\emph{LightGCN}}
\newcommand{\multvae}{\emph{MultVAE}}

\begin{document}
\title{The Impact of Differential Privacy on Recommendation Accuracy and Popularity Bias}
\titlerunning{The Impact of DP on Recommendation Accuracy and Popularity Bias}

\author{Peter Müllner\inst{1}$^{\textrm{(\Letter)}}$\orcidID{0000-0001-6581-1945} \and
Elisabeth Lex\inst{2}\orcidID{0000-0001-5293-2967} \and
Markus Schedl\inst{3,4}\orcidID{0000-0003-1706-3406} \and
Dominik Kowald\inst{1,2}\orcidID{0000-0003-3230-6234}}

\authorrunning{Müllner P., et al.}
%
%
\institute{Know-Center GmbH, Graz, Austria  \and
Graz University of Technology, Graz, Austria \and
Johannes Kepler University Linz, Linz, Austria \and
Linz Institute of Technology, Linz, Austria \\
\email{\{pmuellner, dkowald\}@know-center.at} \\
\email{elisabeth.lex@tugraz.at} \\
\email{markus.schedl@jku.at}}
%
\maketitle              
\begin{abstract}
Collaborative filtering-based recommender systems leverage vast amounts of behavioral user data, which poses severe privacy risks. 
Thus, often random noise is added to the data to ensure Differential Privacy (DP).
However, to date, it is not well understood in which ways this impacts personalized recommendations.
In this work, we study how DP affects recommendation accuracy and popularity bias when applied to the training data of state-of-the-art recommendation models.
Our findings are three-fold:
First, we observe that nearly all users' recommendations change when DP is applied.
Second, recommendation accuracy drops substantially while recommended item popularity experiences a sharp increase, 
suggesting that popularity bias worsens.
Finally, we find that DP exacerbates popularity bias more severely for users who prefer unpopular items than for users who prefer popular items.

\end{abstract}
\section{Introduction}
Modern collaborative filtering-based recommender systems aim to generate personalized recommendations that cater to the specific preferences of each individual user. 
Such recommender systems need to provide recommendations of high accuracy and must ensure that the recommendations do not exhibit popularity bias, i.e., overestimate the relevance of popular items.
For this, vast amounts of user data need to be processed, which exposes the users to many severe privacy risks~\cite{beigi2020survey,shyong2006you,zhang2021membership,calandrino2011you}, e.g., the disclosure of rating data~\cite{calandrino2011you} or the inference of sensitive user attributes~\cite{weinsberg2012blurme,10.1145/3477495.3531820}.
Thus, besides recommendation accuracy and popularity bias, user privacy is another important aspect of recommender systems research. 
Hence, it is critical to leverage privacy-preserving techniques such as \emph{Differential Privacy (DP)}~\cite{dwork2008differential} to devise privacy-aware recommender systems.

Many mechanisms utilized to establish DP include the injection of random noise into the users' interaction data, which typically decreases the overall recommendation accuracy~\cite{berkovsky2012impact,zhu2013differential}.
For recommender systems, some widely used mechanisms are the Gaussian or Laplacian Input Perturbation~\cite{friedman2016differential,dwork2008differential}, Plausible Deniability~\cite{reuseknn,dwork2014algorithmic}, or the 1-Bit mechanism~\cite{ding2017collecting,chen2020practical}.
In particular, the 1-Bit mechanism is a natural match to the binary feedback data prevalent in modern recommender systems.
This mechanism randomly substitutes parts of the positive feedback data with negative or missing feedback data, and then, this modified data is used to train the recommendation model. 
Specifically, the amount of positive feedback data that is randomly substituted depends on the privacy budget $\epsilon$, i.e., a hyperparameter that controls how much random noise is incorporated into the recommendation process and what level of DP is achieved.

However, how DP impacts personalized recommendations is not well understood. 
Specifically, it remains unclear whether DP impacts the recommendations of all users, or just some users, and research on the connection between the $\epsilon$ value and the drop in recommendation accuracy is scarce.
Also, how DP and the $\epsilon$ value impact the item popularity distribution in the respective recommendation lists and thus, popularity bias, is an open research topic.
To shed light on these issues, in this work, we address the following three research questions:
\begin{itemize}
    \item \emph{How many users are impacted by DP? (RQ1)}
    \item \emph{How does the privacy budget $\epsilon$ influence the accuracy drop? (RQ2)}
    \item \emph{In which ways does DP impact popularity bias? (RQ3)}
    \begin{itemize}
        \item[a.] \emph{How does DP impact the popularity distribution of the recommendations?}
        \item[b.] \emph{How does DP impact popularity bias for different user groups?}
    \end{itemize}
\end{itemize}
Accordingly, we perform experiments with a neural matrix-factorization model (i.e., \enmf~\cite{chen2020efficient}), a graph convolution network model (i.e., \lightgcn~\cite{he2020lightgcn}), and a variational autoencoder model (i.e., \multvae~\cite{liang2018variational}), and use datasets from the movie (i.e., \movielens~\cite{harper2015movielens}), music streaming (i.e., \lfm~\cite{kowald2020unfairness}), and online retail domain (i.e., \amazon~\cite{ni2019justifying}).
Plus, we test various $\epsilon$ values to cater for different levels of privacy.

Our results show that nearly all users are impacted by DP, i.e., their recommendations are different from those generated without DP.
Plus, this difference increases when $\epsilon$ becomes smaller (\emph{RQ1}).
With respect to recommendation accuracy, we find that DP leads to a substantial drop, which is most severe for small $\epsilon$ values. 
This highlights the trade-off between recommendation accuracy and privacy (\emph{RQ2}).
Similarly, we present strong evidence that DP increases popularity bias, in particular, when $\epsilon$ becomes smaller.
This underlines an important trade-off between popularity bias and privacy.
Moreover, we identify a ``the poor get poorer'' effect: DP increases popularity bias, especially for users that are already prone to strong popularity bias without DP, i.e., users that prefer unpopular items (cf.~the unfairness of popularity bias~\cite{abdollahpouri2019unfairness,kowald2020unfairness}).

Overall, this work extends existing research on the trade-off between recommendation accuracy and privacy, and contributes novel insights on the connection between DP and popularity bias.



\section{Related Work}
Several previous works~\cite{shyong2006you,calandrino2011you,beigi2020survey,zhang2021membership} identified many critical privacy risks for users in collaborative filtering-based recommender systems.
For example, through the recommendations, the recommender system could leak user data to malicious parties~\cite{calandrino2011you,xin2023user,hashemi2022data}, or an adversary could infer sensitive attributes of the user, e.g., gender~\cite{weinsberg2012blurme,zhang2022comprehensive,10.1145/3477495.3531820}.
To address these privacy risks, privacy-enhancing techniques, such as \emph{Federated Learning}~\cite{lin2020meta,mcmahan2017communication}, \emph{Homomorphic Encryption}~\cite{gentry2009fully,kim2016efficient}, or \emph{Differential Privacy (DP)}~\cite{10.1145/3555374,reuseknn,dwork2008differential} need to be incorporated into the recommender system.
However, Homomorphic Encryption has high computational complexity, and Federated Learning can still leak sensitive user information~\cite{nasr2019comprehensive,ren2022grnn}.

Therefore, in the past years, DP has emerged as a prominent choice in the recommender systems research community.
However, one important shortcoming of DP is its negative impact on recommendation accuracy: DP typically leads to a substantial accuracy drop, since it incorporates random noise into the recommendation process~\cite{berkovsky2012impact,friedman2015privacy}.
Several works address this trade-off between recommendation accuracy and privacy by applying DP in different ways~\cite{friedman2015privacy,10.3389/fdata.2023.1249997}, e.g., by applying DP only to parts of the dataset~\cite{10.5555/2969033.2969119,reuseknn}, or by carefully tuning the degree of noise~\cite{zhu2013differential}.
In detail, Zhu et al.~\cite{zhu2013differential} monitor how strong the item-to-item similarities would change if a piece of user data was not present.
This way, they can better estimate what minimal level of random noise is necessary to ensure DP, and increase recommendation accuracy over comparable approaches.
In a recommender system, there are typically a few users that are willing to openly share their data and many users that are less inclined to share their data.
Xin and Jaakkola~\cite{10.5555/2969033.2969119} exploit this, and protect only a subset of users with DP, which facilitates recommendation accuracy.
Müllner et al.~\cite{reuseknn} attach to this, and modify the recommendation process of user-based KNN to minimize the number of users to which DP needs to be applied.

Besides privacy, another critical problem of recommender systems is popularity bias, i.e., the recommender system overestimates the relevance of popular items and therefore, popular items are overrepresented in the recommendations~\cite{abdollahpouri2019unfairness}.
This can be regarded as disadvantaged, or ``unfair'' treatment of users that prefer unpopular items, since the recommendations do not match these users as well as users that prefer popular items.
In theory, DP and fairness are closely connected to each other~\cite{dwork2012fairness,zemel2013learning}, since for both, a user's data needs to be hidden from the recommender system, either to preserve privacy, or to prohibit discrimination based on, e.g., age or gender.
In practice, correlations in the dataset can still reveal age or gender and, therefore, lead to unfairness~\cite{ekstrand2018privacy,bagdasaryan2019differential}.
In this vein, several works~\cite{ekstrand2018privacy,agarwal2020trade,yang2023fairness} investigate the trade-off between fairness and privacy. 
For example, Sun et al.~\cite{10.1007/978-3-031-28238-6_13} use user data that is protected with DP to rectify the recommendations to increase fairness.
Similarly, also Yang et al.~\cite{yang2023fairness} use post-processing to optimize for fairness with respect to recommendation accuracy.
They observe that regarding recommendation accuracy, DP can lead to more unfairness; however, they do not address DP's impact on popularity bias.

Despite few existing works, how DP impacts personalized recommendations remains an understudied problem and many research gaps exist.
For example, whether the recommendations of all users are impacted, or how beyond-accuracy objectives, such as reducing popularity bias or increasing diversity, are impacted.
Thus, our work attaches to existing work with respect to studying DP's impact on recommendation accuracy, and in addition, we provide novel insights to DP's impact on the trade-off between privacy and popularity bias.

\section{Method}
In this section, we explain how DP is applied to the user data and then, we present multiple evaluation metrics to quantify DP's impact on recommendation accuracy and popularity bias.
Also, we describe the datasets used in this study, and provide all preprocessing steps.
Finally, we detail the experimental setup including the hyperparameters, recommendation models, and our precise evaluation protocol.
We also provide our source-code to foster reproducibiltity.

\subsection{Differential Privacy for Implicit Feedback}
\label{subsec:dp}
To ensure DP, we use the DP-mechanism from Ding et al.~\cite{ding2017collecting}, which is a natural match to the binary implicit feedback data prevalent in today's recommender systems~\cite{chen2020practical}.
With this mechanism, for positive feedbacks $\mathcal{D^+}$ and negative or missing feedbacks $\mathcal{D^-}$ between users and items, the probability that the feedback $f_{u, i}$ between user $u$ and item $i$ is present in the DP dataset $\mathcal{D^+_{DP}}$ is:

\begin{equation}
    Pr[f_{u, i} \in \mathcal{D}^+_{DP}] = \left\{
    \begin{array}{ll}
        \frac{e^\epsilon}{e^\epsilon + 1} & \textrm{\quad if } f_{u, i} \in \mathcal{D^+}  \\
        \frac{1}{e^\epsilon + 1} & \textrm{\quad if } f_{u, i} \in \mathcal{D^-} \\
    \end{array}
    \right.
    \label{eq:dp}
\end{equation}
where $\epsilon$ is the privacy budget~\cite{dwork2008differential} (i.e., it quantifies how much privacy loss is tolerated; the higher, the less noise is added).
In addition to positive feedback data, also negative or missing feedback data can be randomly added to $\mathcal{D}^+_{DP}$.
However, the recommendation model is unable to identify these feedbacks and assumes that all feedbacks in $\mathcal{D}^+_{DP}$ are positive.
By applying this mechanism to the training data of the recommendation model, the recommendations shall not leak information about the data that has been used in the recommendation process.
For computational efficiency, we follow Chen et al.~\cite{chen2020practical} and randomly sample one negative feedback for each positive feedback.

\subsection{Evaluation Metrics}
\label{subsec:evaluation_metrics}
To identify users that are impacted by DP, we compute the Jaccard distance between a user $u$'s recommendation list $\mathcal{R}(u)$ generated without DP and $u$'s recommendation list $\mathcal{R}_{DP}(u)$ generated with DP applied to the training data. 
For the recommendation lists, we use the common cut-off of $n=10$ items.
Formally, the set of users impacted by DP (i.e., $U_{impacted}$) is given by:
\begin{equation}
    U_{impacted} = \bigg\{ u \in U: 1 -
    \frac{|\mathcal{R}(u) \cap \mathcal{R}_{DP}(u)|}{|\mathcal{R}(u) \cup \mathcal{R}_{DP}(u)|} > 0 
    \bigg\}
    \label{eq:n_users}
\end{equation}
where $U$ is the set of all users.
This means that we consider a user $u$ as impacted, if DP leads to at least one different item in $u$'s recommendation list (cf.~\cite{eskandanian2019power}).

Overall, we quantify to what degree DP impacts recommendation accuracy and popularity bias of a user $u$'s recommendations via measuring the relative change of an evaluation metric $\mu$, when DP is applied (cf.~\cite{lesota2021analyzing,muellner2021robustness}):
\begin{align}
    \text{Relative change } \Delta\mu(u) &= \frac{\mu_{DP}(u) - \mu(u)}{\mu(u)} \\
    \label{eq:change}
    \text{Average relative change }\Delta\mu &= \frac{1}{|U_{impacted}|}\sum_{u \in U_{impacted}} \Delta\mu(u)
\end{align}
where $\mu_{DP}(u)$ is the value of the metric for user $u$ when DP is applied and $\mu(u)$ is the value of the metric without applying DP.
Furthermore, $\Delta \mu$ denotes the average change over all impacted users.

\subsubsection{Recommendation Accuracy.}
To study the impact of DP on recommendation accuracy, we compute the ranking-agnostic \recall~\cite{parra2013recommender} metric.
In this work, we use ranking-agnostic metrics since they fit to the way in which we identify impacted users, i.e., whether any item in the recommendation list changes due to DP, disregarding the ordering of the items within the recommendation list.
We do not additionally include \emph{Precision}, since $\Delta Recall = \Delta Precision$\footnote{The number of recommended relevant items is divided by the number of all relevant items (i.e., \emph{Recall}), or by the length of the recommendation list (i.e., \emph{Precision}).
When DP is applied, $\Delta Recall$ and $\Delta Precision$ only depend on how the number of recommended relevant items changes and therefore, the relative change is the same.}.


\subsubsection{Popularity Bias.}
We evaluate DP's impact on popularity bias via measuring the \emph{Average Recommendation Popularity (ARP)}~\cite{klimashevskaia2022mitigating}, i.e., the average fraction of users that interacted with a recommended item:
\begin{equation}
    ARP(u) = \frac{1}{|\mathcal{R}(u)|} \sum_{i \in \mathcal{R}(u)} \phi(i)
\end{equation}
where $\mathcal{R}(u)$ is the recommendation list of user $u$, and item $i$'s popularity $\phi(i)=|U_i| / |U|$ is the number of users that interacted with~$i$, i.e., $|U_i|$, divided by the number of all users $|U|$.
Several works suggest~\cite{abdollahpouri2019unfairness,kowald2020unfairness} that users that prefer unpopular items experience more popularity bias than users that prefer popular items.
Thus, we use \emph{Popularity Lift (PopLift)}~\cite{abdollahpouri2020connection} to quantify popularity bias for distinct user groups. 
Specifically, this metric indicates whether the \emph{ARP} matches the average item popularity $\Gamma(\cdot)$ in the average user's profile of user group $G$: 
\begin{equation}
    PopLift(G) = \frac{\sum_{u \in G} ARP(u) - \sum_{u \in G} \Gamma(u)}{\sum_{u \in G} \Gamma(u)}
\end{equation}
We inspect two user groups: users that prefer items of low popularity, i.e., $U_{low}$, and users that prefer items with high popularity, i.e., $U_{high}$. 
We follow Abdollahpouri et al.~\cite{abdollahpouri2019unfairness} and correspondingly define $U_{low}$ as the set of the 20\% of users with the lowest fraction of popular items in their profile, and $U_{high}$ as the set of the 20\% of users whose profiles contain the highest fraction of popular items.
The set of popular items is given by the 20\% of items with the highest item popularity scores $\phi(i)$.
In addition, we test whether there exists a \emph{Disparate Impact}~\cite{mehrabi2021survey} of DP on $U_{low}$ and $U_{high}$.
Therefore, we measure the \emph{Gap}~\cite{melchiorre2021investigating}, i.e., the absolute difference between the $PopLift$ values of the two user groups:
\begin{equation}
    Gap = |PopLift(U_{low}) - PopLift(U_{high})|
\end{equation}


\subsection{Datasets}
\begin{table}[!t]
    \centering
    \caption{Descriptive statistics of the three datasets. \emph{Users} is the number of users, \emph{Items} is the number of items, \emph{Interactions} is the amount of interactions in the dataset, i.e., positive feedback, \emph{Profile Size} is the average number of interactions per user, 
    and \emph{Density} is the inverse sparsity of the dataset in percent.}
    \setlength{\tabcolsep}{5.1pt}	
    \begin{tabular}{l r r r r r}
        \toprule
        Dataset  & Users & Items & Interactions & Profile Size & Density  \\ \midrule
        \movielens  & 6,038 & 3,533 & 575,281 & 95.28 & 2.70\% \\
        \lfm & 2,999 & 78,799 & 348,437 & 116.18 & 0.15\%  \\
        \amazon  & 3,222 & 6,839 & 72,176 & 22.40 & 0.33\%  \\
         \bottomrule
    \end{tabular}
    \label{tab:datasets}
\end{table}

For our experiments, we use three datasets, i.e., \movielens~\cite{harper2015movielens}, \lfm~\cite{kowald2020unfairness}, and \amazon~\cite{ni2019justifying} (see Table~\ref{tab:datasets}).
\movielens\ and \amazon\ comprise rating scores in the range of 1 to 5, whereas \lfm\ comprises listening events between users and music artists~\cite{lex2020modeling,schedl2021listener}.
For consistency and comparability, we follow~\cite{schedl2017distance} and sum the listening events per artist, followed by scaling the resulting scores to the range of 1 to 5.
For \movielens\ and \lfm, we perform 20-core user pruning, followed by discarding scores below the respective mean value, i.e., 3.58 for \movielens\ and 1.13 for \lfm.
We follow common practice~\cite{sun2020we}, and regard all scores below this threshold, as well as missing scores, as negative feedback.
For \amazon, we additionally 5-core item pruning before filtering the scores according to a threshold of 4.24.

\subsection{Evaluation Protocol}
We split each user's data into 60\% training data used for model training, 20\% validation data used for hyperparameter tuning, and 20\% test data used for evaluation. 
After hyperparameter tuning (see Section~\ref{subsec:params}), to research the impact of DP on personalized recommendations, we add DP to the training data (see Equation~\ref{eq:dp}) and retrain the recommendation models to calculate the evaluation metrics (see Section~\ref{subsec:evaluation_metrics}). 
Specifically, we retrain each model with five different random seeds and average the evaluation metrics to cope for random fluctuations in the training process.
We repeat this procedure for multiple privacy budget values, i.e., $\epsilon \in \{ 5, 4, 3, 2, 1, 0.1, 0.01 \}$. To foster the reproducibility of our research, we publish our source code in an anonymous repository\footnote{\url{https://github.com/pmuellner/ImpactOfDP}}.

\subsection{Recommendation Models and Parameter Settings}
\label{subsec:params}
To cover different kinds of recommender systems, we experiment with a neural matrix-factorization model, i.e., \enmf~\cite{chen2020efficient}, a graph convolution network model, i.e., \lightgcn~\cite{he2020lightgcn}, and a variational autoencoder model, i.e., \multvae~\cite{liang2018variational}.

\begin{itemize}
    \item \enmf~\cite{chen2020efficient} is an efficient neural matrix-factorization model that does not leverage negative sampling. Instead, a negative weighting scheme is used, which benefits training efficiency and recommendation accuracy.
    \item \lightgcn~\cite{he2020lightgcn} is a lightweight graph convolution network, which, in contrast to more complex approaches, only uses neighborhood aggregation and does not include feature transformations or nonlinear activations. 
    \item \multvae~\cite{liang2018variational} is a variational autoencoder that generates recommendations based on a multinomial likelihood.
    This way, it aims to mimic the generative process of implicit feedback data as prevalent in recommender systems. 
    
\end{itemize}
For model training, we use Adam~\cite{kingmaadam} to optimize the models for 5,000 epochs with a batch size of 4,096, and we employ an early stopping threshold of 50.
We perform grid search for every model-dataset pair and determine the hyperparameters of the model with the highest \recall\ on the validation data (see Table~\ref{tab:parameters}).
Note that hyperparameter tuning is performed on the original training data without DP.
\lightgcn\ requires negative samples and therefore, we sample one negative feedback for each positive feedback uniformly at random.
After a careful inspection, we find that with the given hyperparameters, \lightgcn\ cannot produce personalized recommendations for \amazon.
To solve this, we manually adapt the learning rate to 0.001 and the batch size to 1,024. 
In all experiments, the top $10$ ranked items are recommended to each user.

\begin{table}[!t]
    \centering
    \caption{Model hyperparameters used in our experiments (learning rate $\alpha$, dropout probability $\rho$, embedding dimensionality $d$, negative weight $\omega$, $L_2$ regularization factor $\lambda$, number of propagation layers $n$, number of hidden units $h$).}
    \resizebox{\linewidth}{!}{
    \begin{tabular}{l l l l}
        \toprule
        Model & \multicolumn{1}{c}{\movielens} & \multicolumn{1}{c}{\lfm} & \multicolumn{1}{c}{\amazon} \\ \midrule
        \enmf & $\alpha=0.01, \rho=0.1, d=32, \omega=0.25$ & $\alpha=0.001, \rho=0.25, d=128, \omega=0.25$ & $\alpha=0.001, \rho=0.25, d=64, \omega=0.25$\\
        \lightgcn & $\alpha=0.0001 , d=128 , n=1, \lambda=0.0001$ & $\alpha=0.001, d=128, n=4, \lambda=0.01$ & $\alpha=0.001, d=128, n=2, \lambda=0.001$  \\
        \multvae & $\alpha=0.01, \rho=0.5, d=64, h=800$ & $\alpha=0.001, \rho=0.5, d=128, h=600$ & $\alpha=0.0001, \rho=0.5, d=128, h=600$\\ \bottomrule
    \end{tabular}}
    \label{tab:parameters}
\end{table}

\section{Results \& Discussion}
\label{sec:results}


\begin{table*}[!t]
    \centering
    \caption{Absolute values of the evaluation metrics when no DP is used. This serves as baseline for our results in the remainder of this paper, which measure the relative change of the evaluation metrics when DP is applied. For calculating the metrics, we use all impacted users. }
    \resizebox{\linewidth}{!}{
    \begin{tabular}{l r r r r r r r r r}
    \toprule
    & \multicolumn{3}{c}{\movielens} & \multicolumn{3}{c}{\lfm} & \multicolumn{3}{c}{\thead{\emph{Amazon} \\ \emph{Grocery \& Gourmet}}} \\ 
    \cmidrule(l{0.05cm}r{0.05cm}){2-4}\cmidrule(l{0.05cm}r{0.05cm}){5-7}\cmidrule(l{0.05cm}r{0.05cm}){8-10}
    Model & $Recall\uparrow$ & $ARP\downarrow$ & $PopLift\downarrow$ & $Recall\uparrow$ & $ARP\downarrow$ & $PopLift\downarrow$ & $Recall\uparrow$ & $ARP\downarrow$ & $PopLift\downarrow$ \\ \midrule
    \enmf & {0.1697} & 0.2172 & 0.7084 & {0.0971} & 0.0836 & 1.8816 & {0.0932} & {0.0180} & {0.5143} \\
    \lightgcn & 0.1669 & {0.1958} & {0.5405} & 0.0925 & 0.0800 & 1.7585 & 0.0836 & 0.0259 & 1.1796 \\
    \multvae & 0.1694 & 0.1990 & 0.5657 & 0.0835 & {0.0576} & {0.9864} & 0.0643 & 0.0199 & 0.6734 \\ \bottomrule
    \end{tabular}}
    \label{tab:results}
\end{table*}

In this section, we present our results with respect to the three research questions.
First, we measure for how many users the recommendations differ when DP is applied, and we measure how strong these differences are (\emph{RQ1}).
Second, we detail these differences with respect to the relative change of recommendation accuracy (\emph{RQ2}) and popularity bias (\emph{RQ3a}). 
Plus, we investigate the impact of DP on popularity bias from the perspective of two user groups: users that prefer unpopular items and users that prefer popular items (\emph{RQ3b}).
As a baseline, Table~\ref{tab:results} includes the absolute values of our evaluation metrics without DP.

\subsection{Differences Between Recommendations}
\label{sec:rq1}

\begin{table*}[!t]
    \centering
    \caption{\emph{No.~Users} is the percentage of users that are impacted by DP and \emph{Avg.~J.} is the average Jaccard distance between the recommendations with and without DP. The worst results are given in \textbf{bold}. We find that nearly all users are impacted by DP and that the recommendations substantially differ from those generated without DP (\emph{RQ1}). }
    \resizebox{\linewidth}{!}{
    \begin{tabular}{c l r r r r r r}
    \toprule
    & & \multicolumn{2}{c}{\movielens} & \multicolumn{2}{c}{\lfm} & \multicolumn{2}{c}{\thead{\emph{Amazon} \\ \emph{Grocery \& Gourmet}}} \\ 
    \cmidrule(l{0.05cm}r{0.05cm}){3-4}\cmidrule(l{0.05cm}r{0.05cm}){5-6}\cmidrule(l{0.05cm}r{0.05cm}){7-8}
    $\epsilon$ & Model & No.~Users $\downarrow$ & Avg.~J. $\downarrow$ & No.~Users $\downarrow$ & Avg.~J. $\downarrow$. & No.~Users $\downarrow$ & Avg.~J. $\downarrow$ \\ \midrule
    \multirow{3}{*}{\rotatebox[origin=c]{0}{5} }
    & \enmf & 99.41\% & 0.5118 & 98.06\% & 0.4988 & 99.96\% & 0.7872  \\ 
    & \lightgcn & 97.40\% & 0.4207 & 99.14\% & 0.5112 & 99.94\% & 0.7382 \\ 
    & \multvae & 99.71\% & 0.5903 & 99.68\% & 0.6983 & \textbf{100.00\%} & 0.9204 \\ \midrule
    \multirow{3}{*}{\rotatebox[origin=c]{0}{2} }
    & \enmf & 99.85\% & 0.5974 & 99.64\% & 0.5757 & \textbf{100.00\%} & 0.8620 \\ 
    & \lightgcn & 99.86\% & 0.6252 & 99.92\% & 0.6518 & 99.99\% & 0.8132\\ 
    & \multvae & 99.93\% & 0.6828 & \textbf{100.00\%} & 0.7950 & \textbf{100.00\%} & 0.9447\\ \midrule
    \multirow{3}{*}{\rotatebox[origin=c]{0}{1} }
    & \enmf & 99.99\% & 0.7006 & 99.95\% & 0.6858 & \textbf{100.00\%} & 0.9253\\ 
    & \lightgcn & \textbf{99.99\%} & 0.7352 & 99.99\% & 0.7464 & \textbf{100.00\%} & 0.8775 \\ 
    & \multvae & \textbf{100.00\%} & 0.7592 & \textbf{100.00\%} & 0.8408 & \textbf{100.00\%} & 0.9567 \\ \midrule
    \multirow{3}{*}{\rotatebox[origin=c]{0}{0.1} }
    & \enmf & \textbf{100.00\%} & \textbf{0.8183} & \textbf{100.00\%} & \textbf{0.8058} & \textbf{100.00\%} & \textbf{0.9743} \\ 
    & \lightgcn & \textbf{99.99\%} & \textbf{0.8300} & \textbf{100.00\%} & \textbf{0.8490} & \textbf{100.00\%} & \textbf{0.9360}\\ 
    & \multvae & \textbf{100.00\%} & \textbf{0.8447} & \textbf{100.00\%} & \textbf{0.9250} & \textbf{100.00\%} & \textbf{0.9635}\\ \bottomrule
    \end{tabular}
    }
    \label{tab:influence}
\end{table*}

First, we approach \emph{RQ1} and quantify how many users are impacted by DP (see Table~\ref{tab:influence}).
We find that for all datasets, recommendation models, and $\epsilon$ values, DP impacts nearly all users, i.e., different items are recommended than without DP.
For these impacted users, the average difference, i.e., the Jaccard distance between the recommendations with and without DP, lies above 0.5.
Thus, on average, more than every second item in the recommendation list would not have been recommended without DP.
Overall, the impact of DP increases when $\epsilon$ becomes smaller, i.e., when more noise is added to the training data of the recommendation models.
Specifically, for $\epsilon=0.1$ and across all recommendation models and datasets, more than 99.99\% of users are impacted by DP, and the average Jaccard distance lies between 0.8058 and 0.9743.

\emph{This gives strong evidence that DP fundamentally impacts the generated recommendations for nearly all users (RQ1).}


\subsection{Impact on Recommendation Accuracy}
\label{sec:accuracy}

\begin{figure}[!t]
    \centering
    \includegraphics[width=\linewidth]{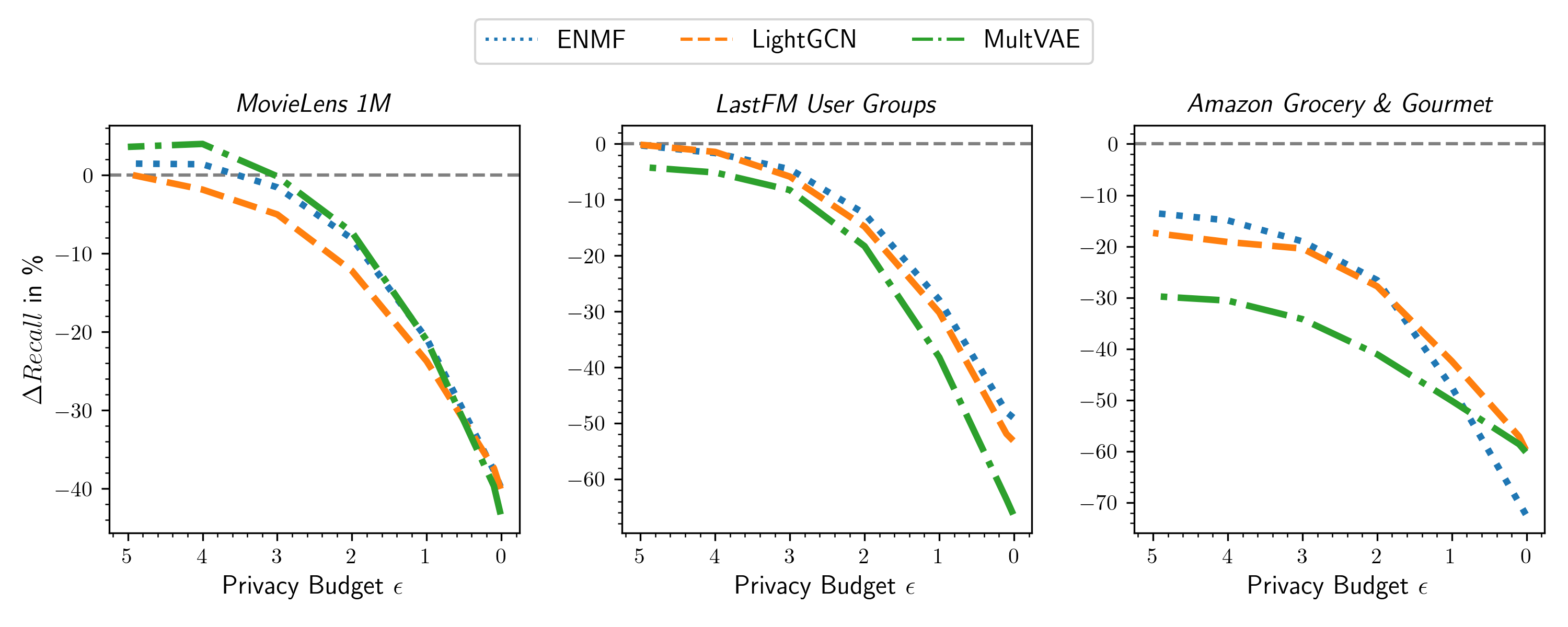}
    \caption{DP's impact on recommendation accuracy as measured by $\Delta$\recall. DP leads to a severe drop in recommendation accuracy. In particular, this drop becomes more serious for small $\epsilon$ values that provide a high level of privacy. This corresponds to the well-known accuracy-privacy trade-off (\emph{RQ2}).}
    \label{fig:delta_recall}
\end{figure}

Next, we build on our results from \emph{RQ1}, and study how DP's impact on the recommendation lists affects recommendation accuracy (\emph{RQ2}).
We find that DP leads to a substantial drop in recommendation accuracy, as measured by \recall\ (see Figure~\ref{fig:delta_recall}).
In contrast to \movielens\ and \lfm, the recommendation accuracy for \amazon\ already drops in case $\epsilon=5$, which is possibly due to DP being applied to the (on average) small user profiles in this dataset (see Table~\ref{tab:datasets}).
In case of \enmf\ and \multvae\ on \movielens, the recommendation accuracy increases slightly for large $\epsilon$ values, which can be possibly due to the fact that the noise introduced by DP acts as Tikhonov regularization for the model~\cite{bishop1995training}.
However, when more noise is added, i.e., $\epsilon < 3$, the recommendation accuracy for these models and dataset drops as well.
Overall, the drop in recommendation accuracy gets worse when $\epsilon$ becomes smaller.
Specifically, for $\epsilon=0.1$ and across all recommendation models, the recommendation accuracy drops by at least 37.39\% (\movielens), 48.00\% (\lfm), or 57.10\% (\amazon).
Since lower $\epsilon$ values lead to higher levels of privacy, this corresponds to the well-known trade-off between recommendation accuracy and privacy~\cite{berkovsky2012impact,zhu2013differential}.

\emph{In summary, DP leads to a substantial drop in recommendation accuracy, and this drop becomes more severe for smaller $\epsilon$ values (RQ2).}

\subsection{Impact on Popularity Bias}
\label{sec:popularity_bias}

\begin{figure}[!t]
    \centering
    \subfloat[t][DP's impact on $ARP$. 
    ]{\includegraphics[width=1\linewidth]{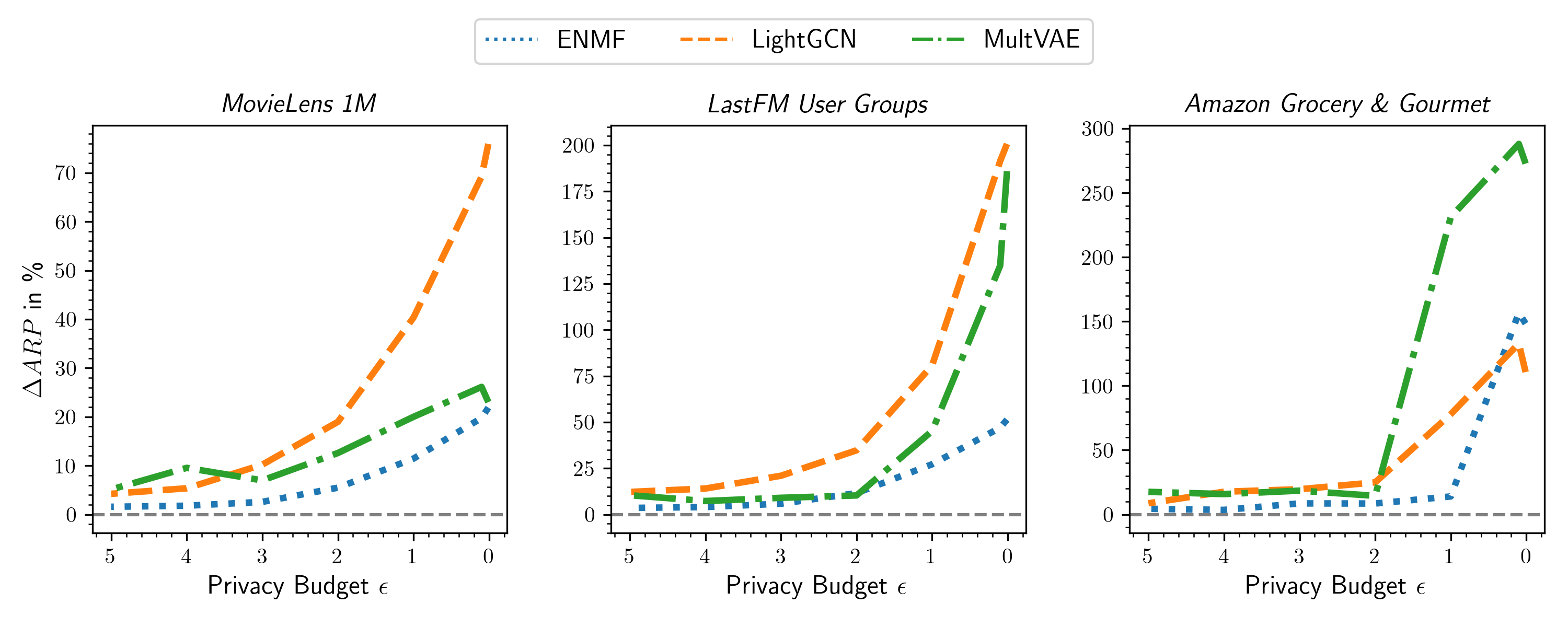}\label{fig:delta_arp}} \\ \vspace{.5cm}
    \subfloat[t][DP's impact on $PopLift$. 
    ]{\includegraphics[width=1\linewidth]{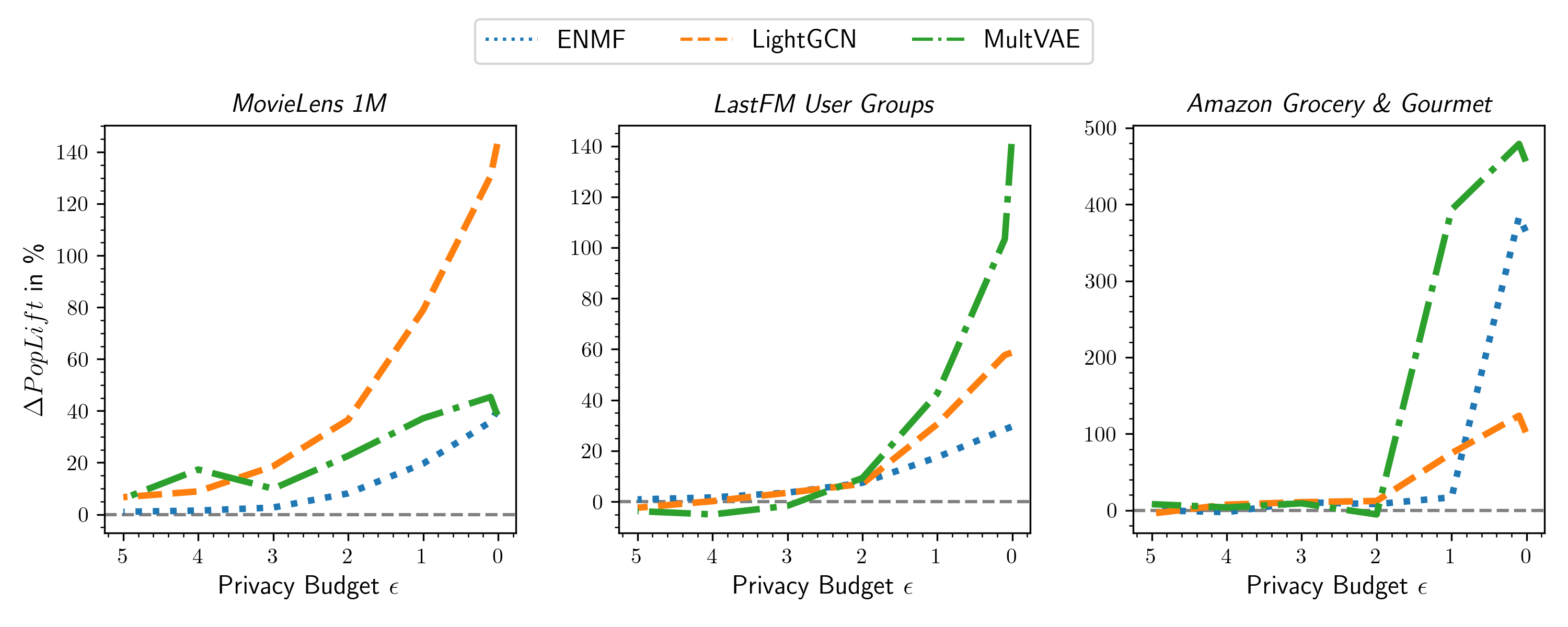}\label{fig:delta_pl}}
    \caption{DP's impact on popularity bias as measured by $\Delta ARP$ and $\Delta PopLift$.
    We find that DP increases \emph{ARP}, which becomes more severe the smaller the $\epsilon$ value is (see Figure~\ref{fig:delta_arp}). Plus, the recommendation popularity mismatches the item popularity distribution in the user profiles (see Figure~\ref{fig:delta_pl}). 
    Overall, this gives strong evidence that DP makes popularity bias worse (\emph{RQ3}).}
    \label{fig:popbias}
\end{figure}

In this section, we study how DP impacts popularity bias (\emph{RQ3}). 
First, in Figure~\ref{fig:popbias}, we monitor how DP impacts the average recommendation popularity (\emph{ARP}) and the popularity lift (\emph{PopLift}).
Then, we investigate DP's impact on popularity bias from the perspective of two user groups: users that prefer unpopular items and users that prefer popular items.


\subsubsection{Impact on Recommendation Popularity.}
We find that DP leads to a substantial increase with respect to \emph{ARP} (see Figure~\ref{fig:delta_arp}).
Specifically, the increase in \emph{ARP} gets worse, when $\epsilon$ becomes smaller.
For example, for $\epsilon=0.1$ and across all recommendation models, \emph{ARP} increases by at least 19.75\% (\movielens), 47.00\% (\lfm), or 132.85\% (\amazon).
We investigate these differences in more detail, and find that the increase is especially high for datasets, for which the baseline value without DP is small (see Table~\ref{tab:results}).
This means that without DP, also items of low popularity are recommended, which are typically hard to recommend (cf. the item cold-start problem~\cite{saveski2014item}).
With the noise introduced by DP, these items are even harder to recommend, which increases the \emph{ARP} value.
Thus, more popular items are recommended as $\epsilon$ becomes smaller, which indicates a trade-off between privacy and popularity bias.
In adddition to \emph{ARP}, we also use \emph{PopLift} to quantify popularity bias, since it relates \emph{ARP} to a user's preference for popular items (see Figure~\ref{fig:delta_pl}).
As in case of \emph{ARP}, also \emph{PopLift} increases when the $\epsilon$ value becomes smaller, i.e., the popularity of the recommended items increasingly mismatches the item popularity distribution in the users' profiles.
Specifically, for $\epsilon=0.1$ and across all recommendation models, \emph{PopLift} increases by at least 36.16\% (\movielens), 28.49\% (\lfm), or 128.38\% (\amazon).
This means that as $\epsilon$ becomes smaller, there is an increasing mismatch between the recommendation popularity and the item popularity distribution of the users.


\emph{Therefore, DP makes the recommendations more biased towards popular items, which strongly overestimates the users' preferences for popular items.
This underlines the important trade-off between privacy and popularity bias (RQ3a).}


\subsubsection{Disparate Impact on User Groups.}

\begin{table*}[!t]
    \centering
    \caption{The absolute \emph{PopLift} values for users that prefer unpopular items ($U_{low}$) and users that prefer popular items ($U_{high}$), and the \emph{Gap}, i.e., the absolute difference, between both. The worst results are given in \textbf{bold}. Popularity bias increases for both user groups with decreasing $\epsilon$, but as \emph{Gap} suggests, popularity bias increases especially for users that prefer unpopular, niche items (\emph{RQ3b}).}
    \resizebox{\linewidth}{!}{
    \begin{tabular}{c l r r r r r r}
    \toprule
    & & \multicolumn{2}{c}{\movielens} & \multicolumn{2}{c}{\lfm} & \multicolumn{2}{c}{\thead{\emph{Amazon} \\ \emph{Grocery \& Gourmet}}} \\ 
    \cmidrule(l{0.05cm}r{0.05cm}){3-4}\cmidrule(l{0.05cm}r{0.05cm}){5-6}\cmidrule(l{0.05cm}r{0.05cm}){7-8}
    & & 
    $PopLift \downarrow$ & $Gap \downarrow$ & $PopLift \downarrow$ & $Gap \downarrow$ & $PopLift \downarrow$ & $Gap \downarrow$   \\ 
    \cmidrule(l{0.05cm}r{0.05cm}){3-3}\cmidrule(l{0.05cm}r{0.05cm}){5-5}\cmidrule(l{0.05cm}r{0.05cm}){7-7}
    $\epsilon$ & Method  & $U_{low}/U_{high}$ & & $U_{low}/U_{high}$ & & $U_{low}/U_{high}$ \\ \midrule
    \multirow{3}{*}{\shortstack{\emph{No DP}}}
    & \enmf & 1.0923/0.4800 & 0.6124 & 4.1028/1.1578 & 2.9450 & 1.4079/0.0845 & 1.3235\\
    & \lightgcn & 0.4225/0.5296 & 0.1072 & 2.7848/1.3273 & 1.4576 & 1.8253/0.7109 & 1.1144 \\
    & \multvae & 0.6247/0.4901 & 0.1347 & 0.7441/0.9092 & 0.1651 & 1.3910/0.2259 & 1.1650\\ \midrule \midrule
    \multirow{3}{*}{\rotatebox[origin=c]{0}{5} }
    & \enmf & 1.0940/0.4903 & 0.6037 & 4.0972/1.1629 & 2.9343 & 1.4043/0.0896 & 1.3147\\
    & \lightgcn & 0.4625/0.5566 & 0.0940 & 2.7952/1.2790 & 1.5162 & 1.7539/0.6766 & 1.0773\\
    & \multvae & 0.6538/0.5227 & 0.1311 & 0.7244/0.8928 & 0.1685 & 1.4372/0.2783 & 1.1589\\ \midrule
    \multirow{3}{*}{\rotatebox[origin=c]{0}{2} }
    & \enmf & 1.2088/0.5147 & 0.6941 & 4.5492/1.2334 & 3.3158 & 1.4977/0.1380 & 1.3597\\
    & \lightgcn & 0.7447/0.6206 & 0.1241 & 3.1516/1.2894 & 1.8623 & 2.0951/0.7736 & 1.3215\\
    & \multvae & 0.8409/0.5814 & 0.2595 & 0.1894/1.0524 & 0.8629 & 1.4175/0.2014 & 1.2161\\ \midrule
    \multirow{3}{*}{\rotatebox[origin=c]{0}{1} }
    & \enmf & 1.3658/0.5612 & 0.8046 & 5.2311/1.3309 & 3.9001 & 1.5723/0.1517 & 1.4206\\
    & \lightgcn & 1.1633/0.7265 & 0.4368 & 3.8118/1.5267 & 2.2851 & 3.1031/1.2728 & 1.8303 \\
    & \multvae & 1.0044/0.6233 & 0.3811 & 0.3395/1.3139 & \textbf{0.9744} & 6.0433/2.0317 & 4.0117 \\ \midrule
    \multirow{3}{*}{\rotatebox[origin=c]{0}{0.1} }
    & \enmf & \textbf{1.5276}/\textbf{0.6445} & \textbf{0.8831} & \textbf{5.7217}/\textbf{1.4448} & \textbf{4.2769} & \textbf{4.9652}/\textbf{1.2375} & \textbf{3.7277} \\
    & \lightgcn & \textbf{1.7767}/\textbf{0.8415} & \textbf{0.9352} & \textbf{5.7233}/\textbf{1.6460} & \textbf{4.0773} & \textbf{4.5163}/\textbf{1.5600} & \textbf{2.9563}\\
    & \multvae & \textbf{1.1595}/\textbf{0.6370} & \textbf{0.5225} & \textbf{1.0760}/\textbf{1.7873} & 0.7113 & \textbf{7.5308}/\textbf{2.3216} & \textbf{5.2092}\\ \bottomrule
    \end{tabular}
    }
    \label{tab:pl}
\end{table*}

Building upon our finding that DP makes popularity bias worse, we finally investigate whether the strength of this effect differs between users that prefer popular items (i.e., $U_{high}$) and users that prefer unpopular items (i.e, $U_{low}$).
For both user groups, \emph{PopLift} increases for small $\epsilon$ values (see Table~\ref{tab:pl}).
Similarly, also the $Gap$ between both user groups' $PopLift$ values grows when $\epsilon$ becomes smaller, which suggests that there exists a disparate impact of DP (cf.~\cite{yang2023fairness}).
We investigate $Gap$ and $PopLift$ in more detail and find that in general, \emph{PopLift} increases more severely for $U_{low}$ than for $U_{high}$.
This can be regarded as a ``poor get poorer'' effect, since these disadvantaged users already experience strong popularity bias without DP.
However, in case of \multvae\ and \lfm, \emph{PopLift} is higher for $U_{high}$ than for $U_{low}$. 
It is known that for some datasets\footnote{No clear pattern across datasets can be observed~\cite{10.1145/3503252.3531292} and thus, this behavior of \multvae\ needs to be researched in the future.} \multvae\ is able to recommend many unpopular items from the long-tail~\cite{10.1145/3503252.3531292}.
This results in lower \emph{ARP} values than in case of the other datasets, i.e., 0.0184 for $U_{low}$ and 0.0933 for $U_{high}$ (without DP), which especially benefits $U_{low}$.
Therefore, this helps to maintain a low \emph{PopLift} value for $U_{low}$, and may explain why in this specific case, \emph{PopLift} is lower for $U_{low}$ than for $U_{high}$.


\emph{Overall, DP makes popularity bias worse for both user groups, but most severely for users that prefer unpopular items (RQ3b).}

\subsection{Discussion}
Overall, DP impacts nearly all users (\emph{RQ1}) and leads to reduced recommendation accuracy (\emph{RQ2}) and increased popularity bias (\emph{RQ3}).
Plus, especially users that prefer unpopular items experience a sharp increase in popularity bias.

However, the impact of DP strongly depends on the level of privacy that shall be ensured, i.e., the $\epsilon$ value.
This suggests that carefully choosing $\epsilon$ is essential in balancing the trade-off between privacy, accuracy, and popularity bias (\emph{RQ2}, \emph{RQ3}).
Moreover, DP's impact on popularity bias is especially severe for users that prefer unpopular items (\emph{RQ3b}).
Thus, the trade-off between privacy, recommendation accuracy, and popularity bias can differ between groups of user, which underlines that group-specific mitigation strategies may be required.
We hope that our results can inform research in this direction.


\section{Conclusion and Future Work}
In this work, we investigated in which ways Differential Privacy (DP) impacts personalized recommendations.
In experiments with three datasets and three recommendation algorithms, we added DP to the training data of state-of-the-art recommendation models, and found that nearly all users' recommendations change when DP is applied.
Also, for higher levels of privacy, recommendation accuracy drops substantially while popularity bias increases. 
In addition, we detail DP's impact on popularity bias and observe a ``poor get poorer'' effect: DP exacerbates popularity bias more severely for users who already experience strong popularity bias without DP, i.e., users who prefer unpopular items.
Overall, our work further researches the trade-off between recommendation accuracy and privacy and, in addition, provides novel insights on the important trade-off between popularity bias and privacy.

\subsubsection*{Future Work.} In the future, we plan to research how users that are especially disadvantaged by DP, i.e., users that prefer unpopular items, can reach a satisfactory trade-off between recommendation accuracy, popularity bias, and privacy.
Specifically, we aim to test whether popularity bias mitigation strategies can help to prohibit the exacerbation of popularity bias for disadvantaged user groups.
One limitation of this work is that we investigated the impact of DP only on the users, but not on other stakeholders of the recommender system. 
Thus, we plan to investigate the impact of DP also on providers and creators of items~\cite{abdollahpouri2020multistakeholder}.
Additionally, we aim to evaluate DP-based recommendations also in more privacy-sensitive domains such as job recommendations~\cite{lacic2020using}. 





%
%
%

\subsubsection*{Acknowledgments.}
This research is funded by the “DDAI” COMET Module within the COMET — Competence Centers for Excellent Technologies Programme, funded by the Austrian Federal Ministry for Transport, Innovation and Technology (bmvit), the Austrian Federal Ministry for Digital and Economic Affairs (bmdw), the Austrian Research Promotion Agency (FFG), the province of Styria (SFG) and partners from industry and academia. The COMET Programme is managed by FFG.
Moreover, this research received support by the Austrian Science Fund (FWF): DFH-23 and P36413; and by the State of Upper Austria and the Federal Ministry of Education, Science, and Research, through grants LIT-2020-9-SEE-113 and LIT-2021-10-YOU-215. For open access purposes, the author has applied a CC BY public copyright license to any author accepted manuscript version arising from this submission.

\bibliographystyle{splncs04}
\bibliography{mybibliography}

\begin{thebibliography}{10}
\providecommand{\url}[1]{\texttt{#1}}
\providecommand{\urlprefix}{URL }
\providecommand{\doi}[1]{https://doi.org/#1}

\bibitem{abdollahpouri2020multistakeholder}
Abdollahpouri, H., Adomavicius, G., Burke, R., Guy, I., Jannach, D., Kamishima,
  T., Krasnodebski, J., Pizzato, L.: Multistakeholder recommendation: Survey
  and research directions. User Modeling and User-Adapted Interaction
  \textbf{30},  127--158 (2020)

\bibitem{abdollahpouri2019unfairness}
Abdollahpouri, H., Mansoury, M., Burke, R., Mobasher, B.: The unfairness of
  popularity bias in recommendation. Workshop on Recommendation in
  Multi-stakeholder Environments (RMSE), in conjunction with the 13th ACM
  Conference on Recommender Systems (RecSys)  (2019)

\bibitem{abdollahpouri2020connection}
Abdollahpouri, H., Mansoury, M., Burke, R., Mobasher, B.: The connection
  between popularity bias, calibration, and fairness in recommendation. In:
  Proceedings of the 14th ACM Conference on Recommender Systems (RecSys). pp.
  726--731 (2020)

\bibitem{agarwal2020trade}
Agarwal, S.: Trade-offs between fairness, interpretability, and privacy in
  machine learning. Master's thesis, University of Waterloo (2020)

\bibitem{10.1145/3503252.3531292}
Anelli, V.W., Bellog{\'\i}n, A., Di~Noia, T., Jannach, D., Pomo, C.: Top-n
  recommendation algorithms: A quest for the state-of-the-art. In: Proceedings
  of the 30th ACM Conference on User Modeling, Adaptation and Personalization
  (UMAP). pp. 121--131 (2022)

\bibitem{bagdasaryan2019differential}
Bagdasaryan, E., Poursaeed, O., Shmatikov, V.: Differential privacy has
  disparate impact on model accuracy. In: Proceedings of the 33rd International
  Conference on Neural Information Processing Systems (NeurIPS). pp.
  15479--15488 (2019)

\bibitem{beigi2020survey}
Beigi, G., Liu, H.: A survey on privacy in social media: identification,
  mitigation, and applications. ACM Transactions on Data Science (TDS)
  \textbf{1}(1),  1--38 (2020)

\bibitem{berkovsky2012impact}
Berkovsky, S., Kuflik, T., Ricci, F.: The impact of data obfuscation on the
  accuracy of collaborative filtering. Expert Systems with Applications
  \textbf{39}(5),  5033--5042 (2012)

\bibitem{bishop1995training}
Bishop, C.M.: Training with noise is equivalent to tikhonov regularization.
  Neural computation  \textbf{7}(1),  108--116 (1995)

\bibitem{calandrino2011you}
Calandrino, J.A., Kilzer, A., Narayanan, A., Felten, E.W., Shmatikov, V.: ``you
  might also like:'' privacy risks of collaborative filtering. In: 2011 IEEE
  Symposium on Security and Privacy (S\&P). pp. 231--246 (2011)

\bibitem{chen2020practical}
Chen, C., Zhou, J., Wu, B., Fang, W., Wang, L., Qi, Y., Zheng, X.: Practical
  privacy preserving poi recommendation. ACM Transactions on Intelligent
  Systems and Technology (TIST)  \textbf{11}(5),  1--20 (2020)

\bibitem{chen2020efficient}
Chen, C., Zhang, M., Zhang, Y., Liu, Y., Ma, S.: Efficient neural matrix
  factorization without sampling for recommendation. ACM Transactions on
  Information Systems (TOIS)  \textbf{38}(2),  1--28 (2020)

\bibitem{ding2017collecting}
Ding, B., Kulkarni, J., Yekhanin, S.: Collecting telemetry data privately. In:
  Proceedings of the 31st International Conference on Neural Information
  Processing Systems (NeurIPS). pp. 3574--3583 (2017)

\bibitem{dwork2008differential}
Dwork, C.: Differential privacy: A survey of results. In: International
  conference on theory and applications of models of computation (TAMC). pp.
  1--19 (2008)

\bibitem{dwork2012fairness}
Dwork, C., Hardt, M., Pitassi, T., Reingold, O., Zemel, R.: Fairness through
  awareness. In: Proceedings of the 3rd innovations in theoretical computer
  science conference (ITCS). pp. 214--226 (2012)

\bibitem{dwork2014algorithmic}
Dwork, C., Roth, A., et~al.: The algorithmic foundations of differential
  privacy. Now Publishers, Inc. (2014)

\bibitem{ekstrand2018privacy}
Ekstrand, M.D., Joshaghani, R., Mehrpouyan, H.: Privacy for all: Ensuring fair
  and equitable privacy protections. In: Proceedings of ACM Conference on
  Fairness, Accountability, and Transparency (FAccT). pp. 35--47 (2018)

\bibitem{eskandanian2019power}
Eskandanian, F., Sonboli, N., Mobasher, B.: Power of the few: Analyzing the
  impact of influential users in collaborative recommender systems. In:
  Proceedings of the 27th ACM Conference on User Modeling, Adaptation and
  Personalization. pp. 225--233 (2019)

\bibitem{friedman2016differential}
Friedman, A., Berkovsky, S., Kaafar, M.A.: A differential privacy framework for
  matrix factorization recommender systems. User Modeling and User-Adapted
  Interaction (UMUAI)  \textbf{26}(5),  425--458 (2016)

\bibitem{friedman2015privacy}
Friedman, A., Knijnenburg, B.P., Vanhecke, K., Martens, L., Berkovsky, S.:
  Privacy Aspects of Recommender Systems, pp. 649--688. Springer US, Boston, MA
  (2015). \doi{10.1007/978-1-4899-7637-6\_19"}

\bibitem{10.1145/3477495.3531820}
Ganh{\"o}r, C., Penz, D., Rekabsaz, N., Lesota, O., Schedl, M.: Unlearning
  protected user attributes in recommendations with adversarial training. In:
  Proceedings of the 45th International ACM SIGIR Conference on Research and
  Development in Information Retrieval (SIGIR). pp. 2142--2147. Springer (2022)

\bibitem{gentry2009fully}
Gentry, C.: A fully homomorphic encryption scheme. Ph.D. thesis, Stanford
  university (2009)

\bibitem{harper2015movielens}
Harper, F.M., Konstan, J.A.: The movielens datasets: History and context. ACM
  Transactions on Interactive Intelligent Systems (TiiS)  \textbf{5}(4),  1--19
  (2015)

\bibitem{hashemi2022data}
Hashemi, H., Xiong, W., Ke, L., Maeng, K., Annavaram, M., Suh, G.E., Lee,
  H.H.S.: Data leakage via access patterns of sparse features in deep
  learning-based recommendation systems. Workshop on Trustworthy and Socially
  Responsible Machine Learning (TSRML), in conjunction with the 36th Conference
  on Neural Information Processing Systems (NeurIPS)  (2022)

\bibitem{he2020lightgcn}
He, X., Deng, K., Wang, X., Li, Y., Zhang, Y., Wang, M.: Lightgcn: Simplifying
  and powering graph convolution network for recommendation. In: Proceedings of
  the 43rd International ACM SIGIR conference on research and development in
  Information Retrieval (SIGIR). pp. 639--648. Springer (2020)

\bibitem{kim2016efficient}
Kim, S., Kim, J., Koo, D., Kim, Y., Yoon, H., Shin, J.: Efficient
  privacy-preserving matrix factorization via fully homomorphic encryption. In:
  Proceedings of the 11th ACM on Asia conference on computer and communications
  security (ASIACCS). pp. 617--628 (2016)

\bibitem{kingmaadam}
Kingma, D.P., Ba, J.: Adam: {A} method for stochastic optimization. In:
  Proceedings of 3rd International Conference on Learning Representations
  (ICLR) (2015)

\bibitem{klimashevskaia2022mitigating}
Klimashevskaia, A., Elahi, M., Jannach, D., Trattner, C., Skj{\ae}rven, L.:
  Mitigating popularity bias in recommendation: Potential and limits of
  calibration approaches. In: Advances in Information Retrieval: Workshop on
  Algorithmic Bias in Search and Recommendation (BIAS) in conjunction with the
  42nd European Conference on IR Research (ECIR). pp. 82--90. Springer (2022)

\bibitem{kowald2020unfairness}
Kowald, D., Schedl, M., Lex, E.: The unfairness of popularity bias in music
  recommendation: A reproducibility study. In: Advances in Information
  Retrieval: 42nd European Conference on IR Research (ECIR). pp. 35--42.
  Springer (2020)

\bibitem{lacic2020using}
Lacic, E., Reiter-Haas, M., Kowald, D., Reddy~Dareddy, M., Cho, J., Lex, E.:
  Using autoencoders for session-based job recommendations. User Modeling and
  User-Adapted Interaction  \textbf{30},  617--658 (2020)

\bibitem{lesota2021analyzing}
Lesota, O., Melchiorre, A., Rekabsaz, N., Brandl, S., Kowald, D., Lex, E.,
  Schedl, M.: Analyzing item popularity bias of music recommender systems: are
  different genders equally affected? In: Proceedings of the 15th ACM
  Conference on Recommender Systems (RecSys). pp. 601--606 (2021)

\bibitem{lex2020modeling}
Lex, E., Kowald, D., Schedl, M.: Modeling popularity and temporal drift of
  music genre preferences. Transactions of the International Society for Music
  Information Retrieval  \textbf{3}(1) (2020)

\bibitem{liang2018variational}
Liang, D., Krishnan, R.G., Hoffman, M.D., Jebara, T.: Variational autoencoders
  for collaborative filtering. In: Proceedings of the World Wide Web Conference
  (TheWebConf). pp. 689--698 (2018)

\bibitem{lin2020meta}
Lin, Y., Ren, P., Chen, Z., Ren, Z., Yu, D., Ma, J., Rijke, M.d., Cheng, X.:
  Meta matrix factorization for federated rating predictions. In: Proceedings
  of the 43rd International ACM SIGIR Conference on Research and Development in
  Information Retrieval (SIGIR). pp. 981--990. Springer (2020)

\bibitem{10.1145/3555374}
Long, J., Chen, T., Nguyen, Q.V.H., Yin, H.: Decentralized collaborative
  learning framework for next poi recommendation. ACM Trans. Inf. Syst.
  \textbf{41}(3) (2023). \doi{10.1145/3555374}

\bibitem{mcmahan2017communication}
McMahan, B., Moore, E., Ramage, D., Hampson, S., y~Arcas, B.A.:
  Communication-efficient learning of deep networks from decentralized data.
  In: Proceedings of the 20th International Conference on Artificial
  Intelligence and Statistics (AISTATS). pp. 1273--1282 (2017)

\bibitem{mehrabi2021survey}
Mehrabi, N., Morstatter, F., Saxena, N., Lerman, K., Galstyan, A.: A survey on
  bias and fairness in machine learning. ACM computing surveys (CSUR)
  \textbf{54}(6),  1--35 (2021)

\bibitem{melchiorre2021investigating}
Melchiorre, A.B., Rekabsaz, N., Parada-Cabaleiro, E., Brandl, S., Lesota, O.,
  Schedl, M.: Investigating gender fairness of recommendation algorithms in the
  music domain. Information Processing \& Management (IP\&P)  \textbf{58}(5),
  102666 (2021)

\bibitem{reuseknn}
M\"{u}llner, P., Lex, E., Schedl, M., Kowald, D.: Reuseknn: Neighborhood reuse
  for differentially-private knn-based recommendations. ACM Trans. Intell.
  Syst. Technol.  (2023). \doi{10.1145/3608481}

\bibitem{muellner2021robustness}
Müllner, P., Kowald, D., Lex, E.: Robustness of meta matrix factorization
  against strict privacy constraints. In: Advances in Information Retrieval:
  43rd European Conference on IR Research (ECIR). pp. 107--119. Springer (2021)

\bibitem{10.3389/fdata.2023.1249997}
Müllner, P., Lex, E., Schedl, M., Kowald, D.: Differential privacy in
  collaborative filtering recommender systems: a review. Frontiers in Big Data
  \textbf{6} (2023). \doi{10.3389/fdata.2023.1249997}

\bibitem{nasr2019comprehensive}
Nasr, M., Shokri, R., Houmansadr, A.: Comprehensive privacy analysis of deep
  learning: Passive and active white-box inference attacks against centralized
  and federated learning. In: Proceedings of the IEEE Symposium on Security and
  Privacy (S\&P). pp. 739--753 (2019)

\bibitem{ni2019justifying}
Ni, J., Li, J., McAuley, J.: Justifying recommendations using distantly-labeled
  reviews and fine-grained aspects. In: Proceedings of the conference on
  empirical methods in natural language processing and the 9th international
  joint conference on natural language processing (EMNLP-IJCNLP). pp. 188--197
  (2019)

\bibitem{parra2013recommender}
Parra, D., Sahebi, S.: Recommender systems: Sources of knowledge and evaluation
  metrics. In: Advanced Techniques in Web Intelligence-2: Web User Browsing
  Behaviour and Preference Analysis, pp. 149--175. Springer (2013)

\bibitem{ren2022grnn}
Ren, H., Deng, J., Xie, X.: Grnn: Generative regression neural network—a data
  leakage attack for federated learning. ACM Transactions on Intelligent
  Systems and Technology (TIST)  \textbf{13}(4),  1--24 (2022)

\bibitem{saveski2014item}
Saveski, M., Mantrach, A.: Item cold-start recommendations: learning local
  collective embeddings. In: Proceedings of the 8th ACM Conference on
  Recommender systems (RecSys). pp. 89--96 (2014)

\bibitem{schedl2017distance}
Schedl, M., Bauer, C.: Distance-and rank-based music mainstreaminess
  measurement. In: Adjunct publication of the 25th conference on user modeling,
  adaptation and personalization (UMAP): Workshop on Surprise, Opposition, and
  Obstruction in Adaptive and Personalized Systems (SOAP). pp. 364--367 (2017)

\bibitem{schedl2021listener}
Schedl, M., Bauer, C., Reisinger, W., Kowald, D., Lex, E.: Listener modeling
  and context-aware music recommendation based on country archetypes. Frontiers
  in Artificial Intelligence  \textbf{3},  508725 (2021)

\bibitem{shyong2006you}
Shyong, K., Frankowski, D., Riedl, J., et~al.: Do you trust your
  recommendations? an exploration of security and privacy issues in recommender
  systems. In: International Conference on Emerging Trends in Information and
  Communication Security (ETRICS ). pp. 14--29. Springer (2006)

\bibitem{10.1007/978-3-031-28238-6_13}
Sun, J.A., Pentyala, S., Cock, M.D., Farnadi, G.: Privacy-preserving fair item
  ranking. In: European Conference on Information Retrieval (ECIR). pp.
  188--203. Springer (2023)

\bibitem{sun2020we}
Sun, Z., Yu, D., Fang, H., Yang, J., Qu, X., Zhang, J., Geng, C.: Are we
  evaluating rigorously? benchmarking recommendation for reproducible
  evaluation and fair comparison. In: Proceedings of the 14th ACM Conference on
  Recommender Systems (RecSys). pp. 23--32 (2020)

\bibitem{weinsberg2012blurme}
Weinsberg, U., Bhagat, S., Ioannidis, S., Taft, N.: Blurme: Inferring and
  obfuscating user gender based on ratings. In: Proceedings of the 6th ACM
  conference on Recommender systems (RecSys). pp. 195--202 (2012)

\bibitem{xin2023user}
Xin, X., Yang, J., Wang, H., Ma, J., Ren, P., Luo, H., Shi, X., Chen, Z., Ren,
  Z.: On the user behavior leakage from recommender system exposure. ACM
  Transactions on Information Systems (TOIS)  \textbf{41}(3),  1--25 (2023)

\bibitem{10.5555/2969033.2969119}
Xin, Y., Jaakkola, T.: Controlling privacy in recommender systems. In:
  Proceedings of the 27th International Conference on Neural Information
  Processing Systems (NeurIPS). pp. 2618--2626. MIT Press, Cambridge, MA, USA
  (2014)

\bibitem{yang2023fairness}
Yang, Z., Ge, Y., Su, C., Wang, D., Zhao, X., Ying, Y.: Fairness-aware
  differentially private collaborative filtering. In: Companion Proceedings of
  the {ACM} Web Conference (TheWebConf). pp. 927--931 (2023)

\bibitem{zemel2013learning}
Zemel, R., Wu, Y., Swersky, K., Pitassi, T., Dwork, C.: Learning fair
  representations. In: International conference on machine learning (ICML). pp.
  325--333 (2013)

\bibitem{zhang2021membership}
Zhang, M., Ren, Z., Wang, Z., Ren, P., Chen, Z., Hu, P., Zhang, Y.: Membership
  inference attacks against recommender systems. In: Proceedings of the ACM
  SIGSAC Conference on Computer and Communications Security (CCS). pp. 864--879
  (2021)

\bibitem{zhang2022comprehensive}
Zhang, S., Yin, H.: Comprehensive privacy analysis on federated recommender
  system against attribute inference attacks. IEEE Transactions on Knowledge
  and Data Engineering (TKDE)  (2023)

\bibitem{zhu2013differential}
Zhu, T., Li, G., Ren, Y., Zhou, W., Xiong, P.: Differential privacy for
  neighborhood-based collaborative filtering. In: Proceedings of the IEEE/ACM
  International Conference on Advances in Social Networks Analysis and Mining
  (ASONAM). pp. 752--759 (2013)

\end{thebibliography}
\end{document}